\documentclass[oldversion]{aa}
\usepackage{graphicx}
\usepackage{natbib}
\usepackage{txfonts}
\def\d{d$^{-1}$}

\def\eo{$\epsilon$ Oph}

\begin{document}
   \title{Nonradial $p$-modes in the G9.5 giant $\epsilon$ Oph? \\ Pulsation model fits to MOST\thanks{Based on data from the {\it MOST} satellite, a Canadian Space Agency mission jointly operated by Dynacon, Inc., the University of Toronto Institute of Aerospace Studies, and the University of British Columbia, with assistance from the University of Vienna, Austria.} photometry}

   \author{Thomas Kallinger\inst{1}
   	 \and
          David B. Guenther\inst{2}
          \and
          Jaymie M. Matthews\inst{3}
          \and
          Werner W. Weiss\inst{1} 
          \and \\
          Daniel Huber\inst{1}
          \and
          Rainer Kuschnig\inst{3}
          \and
          Anthony F. J. Moffat\inst{4}
          \and
          Slavek M. Rucinski\inst{5} \\
          and
          Dimitar Sasselov\inst{6}
          }

   \offprints{kallinger@astro.univie.ac.at}

   \institute{Institute for Astronomy (IfA), University of Vienna,
              T\"urkenschanzstrasse 17, A-1180 Vienna
              \and
              Institute for Computational Astrophysics, Department of Astronomy and Physics, Saint Marys University, 
Halifax, NS B3H 3C3, Canada
              \and
              Department of Physics and Astronomy, University of British Columbia, 6224 Agricultural Road, 
Vancouver, BC V6T 1Z1, Canada
		\and
		Observatoire Astronomique du Mont M\'egantic, D\'epartement de Physique, Universit\'e de Montr\'eal,
C. P. 6128, Succursale: Centre-Ville, Montr\'eal, QC H3C 3J7 and Observatoire du mont M\'egantic, Canada
		\and
		Department of Astronomy and Astrophysics, David Dunlap Observatory, University of Toronto, P. O. Box 360, Richmond Hill, ON L4C 4Y6, Canada
		\and
		Harvard-Smithsonian Center for Astrophysics, 60 Garden Street, Cambridge, MA 02138
             }

   \date{Received ; accepted }

   \abstract
   {The G9.5 giant \eo\ shows evidence of radial $p$-mode pulsations in both radial velocity and luminosity.  We re-examine the observed frequencies in the photometry and radial velocities and find a best model fit to 18 of the 21 most significant photometric frequencies. The observed frequencies are matched to both radial and $non$radial modes in the best model fit. The small scatter of the frequencies about the model predicted frequencies indicate that the average lifetimes of the modes could be as long as 10--20\,d. The best fit model itself, constrained $only$ by the observed frequencies, lies within $\pm1\sigma$ of \eo 's position in the HR-diagram and the interferometrically determined radius.}

\authorrunning{Kallinger et al.}
\titlerunning{$Non$radial $p$-modes in the G9.5 $\epsilon$ Oph?}
   \maketitle

\section{Introduction}
In radial velocity measurements from ground, De Ridder et al. (2006) found power excess in the frequency spectrum of $\epsilon$ Oph with a maximum amplitude of about 3.5 ms$^{-1}$ at a frequency of approximately 60  $\mu$Hz 
($\sim$5.2 \d ).  However, the modest duty cycle of the groundbased observations and the resulting \d\ aliases made it difficult for them to choose between possible $p$-mode spacings of $\sim$5 and $\sim$7  $\mu$Hz. Barban et al. (2007) analyzed recent MOST ({\it Microvariability} \& {\it Oscillations of STars}) satellite observations and obtained frequencies, amplitudes, and mode lifetimes for a sequence of radial modes. Their mode lifetimes ($\sim$2.7 d) compared to another giant $\xi$ Hya are consistent with the lifetimes determined from observations (Stello et al. 2006) but are shorter than the lifetimes inferred from theoretical considerations (e.g., Houdek and Gough 2002).

Barban et al. (2007) only considered radial modes in their analysis. Indeed, it has been suggested that only radial $p$-modes should be observable in giants owing in part to the complex frequency spacings of nonradial mixed (or bumped) modes. The theoretical position is unclear. The nonadiabatic models of Dziembowski et al. (2001) show the existence of strongly trapped unstable (STU) nonradial modes in the central regions of giants in between very closely spaced modes of the same degree $l$. The STU modes follow the same pattern of frequency spacing as radial acoustic modes (i.e., $p$-modes). They correspond to mixed $g$-modes/$p$-modes. Furthermore, the authors find more $l$ = 2 STU modes to be unstable than $l$ = 1 STU modes for stars in the luminosity range of $\epsilon$ Oph (see Fig. 4 in Dziembowski et al. 2001). Only in their linear stability analysis, which includes the effects of turbulent pressure based on the mixing-length formalism, do they find predicted amplitudes of the radial modes of the sub-giant star $\alpha$ UMa to be much smaller than observed, with the nonradial mode amplitudes predicted to be even smaller. On the observational side, Hekker et al. (2006) note that the line profile variations of several pulsating red giants (including $\epsilon$ Oph) suggest the existence of nonradial modes. 

If some or all pulsating red giants do have radial and observable nonradial modes with relatively long lifetimes, then these modes will enable asteroseismologists to constrain the deep interiors of red giant stars, as well as set limits on the excitation mechanisms themselves. We are therefore motivated to re--examine the MOST photometry of $\epsilon$ Oph, combined with the radial velocities obtained by De Ridder et al. (2006), to explore the possibility of longer--lived (longer than 3 days) nonradial pulsations.


   \begin{figure}[t]
   \centering
      \includegraphics[width=0.5\textwidth]{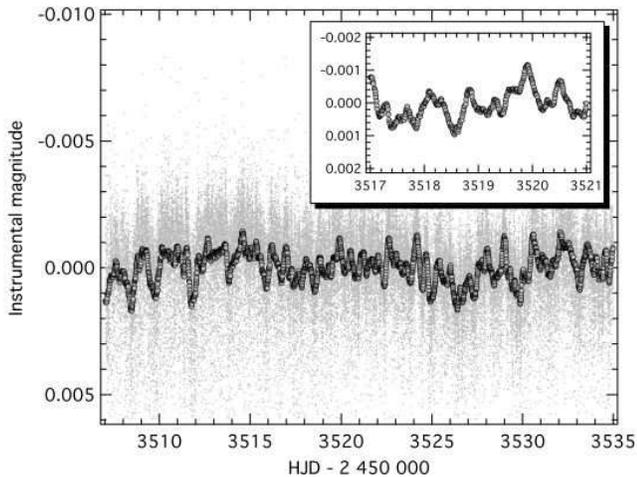}
      \caption{MOST photometry of $\epsilon$ Oph (V=3$^{m}$24). For better visibility, circles are averages of 20 consecutive data points (covering about 10 min each) of the smoothed raw data (running average of 350 exposures); the un-averaged measurements of $\epsilon$ Oph are shown as small grey dots. The inset is an enlargement of 4 day subset of the data.}
         \label{FigLC}
   \end{figure}

\section{Observations}
\subsection{Spectroscopy}
The radial velocity data of $\epsilon$ Oph were provided to us by De Ridder et al. (2006) from their double-site campaign with the fibre--fed echelle spectrographs CORALIE (Swiss 1.2-m Euler telescope at La Silla, Chile) and ELODIE (French 1.93-m telescope at the Observatoire de Haute-Provence).  The spectra were obtained during 53 nights (spanning 75 days) in summer 2003, with S/N of at least 100 in every spectrum at a wavelength of 550 nm. Radial velocities were measured from the spectra using the optimum-weight method relative to a nightly reference spectrum. This suppresses any signal (including signal intrinsic to the star) with periods longer than about 1 day. 

\subsection{Photometry}
MOST houses a CCD photometer fed by a 15-cm Maksutov telescope through a custom broadband optical filter (350 - 750nm).  The satellite's Sun-synchronous 820\,km polar orbit (period = 101.4\,min) enables uninterrupted observations of stars in its Continuous Viewing Zone ($+36^{\circ}$$\geq$$\delta$$\geq$$-18^{\circ}$) for up to 8 weeks.  A pre-launch summary of the mission is given by Walker et al. (2003) and on-orbit science operations are described by Matthews et al. (2004). 

The MOST photometry of $\epsilon$ Oph is described by Barban et al. (2007).  The data cover 28.4 d during 16 May -- 13 June 2005, and were obtained in Fabry Imaging mode, where the telescope pupil, illuminated by the star, is projected onto the CCD by a Fabry microlens as an annulus covering about 1300 pixels. To meet satellite downlink restrictions, the data are downloaded in two formats: resolved CCD subrasters of the stellar and background Fabry images (Science Data Stream 2 or SDS2), and compressed data preprocessed on board the satellite (SDS1). We use the resolved SDS2 data for our analysis here because it allows us to carry out background decorrelation of the Fabry images (Reegen et al. 2006). Our subset of the MOST photometry contains 83,238 measurements with an exposure time of 7.5s, sampled every 30s. 

As in the Barban et al. (2007) analysis, for our re-analysis of $\epsilon$ Oph we reject images obtained during MOST orbital phases of high scattered Earthshine, which for the $\epsilon$ Oph run, occurred when the satellite was passing over the fully illuminated North Pole.  We introduce gaps in the data of about 24 min duration every 101.4\,min (the MOST orbital period). We also reject all 3$\sigma$ outlier frames. In total 24\% of the exposures were rejected yielding an overall duty cycle of 76\%.  The regular gaps per orbit introduce aliases in the spectral window of the data, but these alias peaks are spaced $\sim$14.2\d\ ($\sim$165 $\mu$Hz) apart. Fortunately, this alias spacing, unlike the 1\d\ ground based alias spacing, does not introduce any ambiguities in the frequency identifications in the range where $\epsilon$ Oph exhibits $p$-mode pulsations. 

MOST was not originally designed to obtain differential photometry, but thanks to its stable thermal environment and compact design, the instrument is sensitive to stellar variability at the millimagnitude level (and below) over timescales of up to one week. Only for very long timescales, comparable to the duration of the observations, instrumental drifts became severe and superpose possible intrinsic variations. Stars like the red giant $\epsilon$ Oph have turbulent envelopes which probably introduce longer-term variability than the $p$-modes. Regardless, all long term variations (i.e., longer than a week), whether or not they are intrinsic, are ignored. 

Fig.\,\ref{FigLC} presents the binned light curve of $\epsilon$ Oph, as well as the original unbinned reduced photometry. The data show very complex variations on timescales of hours to days, also seen by Barban et al. (2007), in addition to some slower trends. The clean spectral window of the data, with alias peaks spaced by 14.2\d , shows that the slower trends do not interfere with the $p$-mode identifications. There exists some modulation of stray light at low harmonics of 1\d . This is caused by the fact that the MOST orbit carries the spacecraft over nearly the same location on the Earth (hence, a similar albedo feature) every day.  

The reduced MOST $\epsilon$ Oph photometry used in this paper, as well as the raw data and the Barban et al. (2006) reduction, are available in the MOST Public Data Archive on the Science page at {\it www.astro.ubc.ca/MOST}.


   \begin{figure*}[t]
   \centering
      \includegraphics[width=\textwidth]{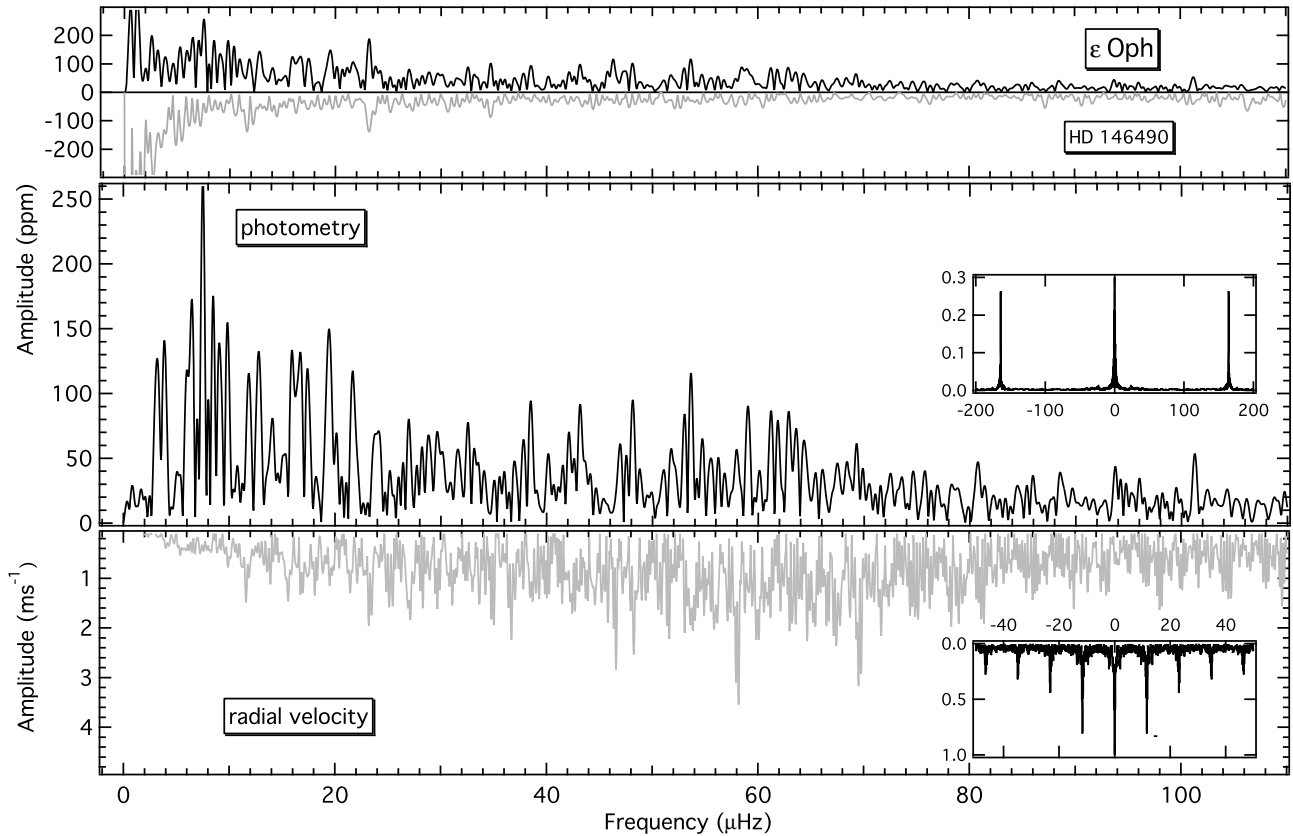}
      \caption{{\it Top panel}: Fourier amplitude spectra of $\epsilon$ Oph, and that of an intrinsically constant MOST guide star, HD 146490 (inverted for easier comparison). Both are shown at the same scale. {\it Middle} and {\it bottom panel}:  Residual amplitude spectra of $\epsilon$ Oph MOST photometry after prewhitening instrumental signal and radial velocity data (De Ridder et al. 2006). The spectral window functions of both data sets are shown in the insets; note the truncated amplitude scale for MOST spectral window.}
         \label{FigFourier}
   \end{figure*}

\section{Frequency Analysis}
The Fourier amplitude spectra of the $\epsilon$ Oph MOST photometry (already prewhitened for instrumental signal; see latter in this section) and the radial velocity (RV) data are plotted in Fig.\,\ref{FigFourier} along with the spectral window functions of both data sets. In the photometric amplitude spectrum, there are no aliases in the frequency range plotted, since the sampling sidelobes fall well outside the $p$-mode frequency range.  There are, however, peaks at 2 and 3\d\ ($\sim$23.1 and $\sim$34.7$\mu$Hz) due to modulation of stray light in the data, as mentioned above. Some of the power at frequencies below 10  $\mu$Hz may be due to intrinsic turbulence in the stellar envelope, and some may be due to slow instrumental drifts but there is no significant leakage of this power to higher frequencies in the $p$-mode range because of the clean spectral window.

The RV amplitude spectrum shows the cycle/day (\d ) alias patterns due to the daily gaps in the groundbased spectroscopy.  The data were subjected to a high-pass filter by De Ridder et al. (2006) so amplitudes at frequencies below $\sim$20$\mu$Hz are suppressed. The RV spectrum has better frequency resolution than the photometric spectrum because of the longer time coverage of the RV data. 

We used the routine {\it SigSpec} (Reegen 2007) to identify significant frequencies in the Discrete Fourier Transform (DFT) spectra of both data sets. {\it SigSpec} employs an exact analytical solution for the probability that a peak of given amplitude could be generated by white noise. Its main advantage over commonly used signal-to-noise ratio estimates is that {\it SigSpec} appropriately incorporates the frequency and phase angle in Fourier space and the time domain sampling. On average, a signal-to-noise ratio of 4, suggested as a reliable significance estimator by Breger et al. (1993), roughly corresponds to a {\it SigSpec} significance value of 5.46. The {\it SigSpec} significance is equal to log$_{10}$ the probability that the peak is not due to white noise. In other words, a signal amplitude of 4 times the noise level would appear by chance at its given frequency in only one of 10$^{5.46}$ cases, assuming white noise.

{\it SigSpec} automatically identifies the peak with the largest significance in the frequency range searched, determines the amplitude and phase associated with this frequency through least-squares sinusoidal fitting, and subtracts that signal from the time series. The residuals are then used in the next {\it SigSpec} significance calculation, and the process is repeated until there are no frequencies remaining with {\it SigSpec} significance above the threshold value set for the search.

In the MOST photometry we find 99 peaks with {\it SigSpec} significances (sig) greater than 4.8  (equivalent to S/N $\sim$ 3.75) from 0 - 350$\mu$Hz ($\sim$30\d ). 17 of the peaks are coincident with the MOST satellite orbital frequency ($\sim$165$\mu$Hz or 14.2\d ) and its first overtone as well as with sidelobes produced by the 1 cycle/day and long term modulation of the stray light amplitude.  We compare the remaining 82 frequencies (all below $\sim$120$\mu$Hz or $\sim$10\d ) with frequencies in the background signal and with frequencies in the light curve of the simultaniously observed guide star HD 146490 (V = 7$^{m}$2, sp. A2).  (MOST guide star photometry is collected from defocused star images in subrasters of 40$\times$40 pixels. Although, not suitable for differential comparisons with the Fabry Imaging photometry, the guide star photometry can be used to help identify instrumental or environmental frequencies in the data.)  The background time series was extracted from the raw $\epsilon$ Oph Fabry Image data using the average intensity of pixels well outside the annulus illuminated by stellar light that contains only signal from the sky background along with the bias and dark current levels of the CCD. {\it SigSpec} identified 34 peaks, from 0 to 120$\mu$Hz, with sig $\geq$ 4.8 in the background signal, and 14 peaks in the guide star, HD 146490, time series. 

We reject all frequencies that occur (within the uncertainties) in both the $\epsilon$ Oph photometry and either the background or HD 146490 time series, even if the amplitudes in the latter data sets are small. The frequency uncertainties are estimated according to Kallinger et al. (2007b) who define an upper limit for the frequency error of a least-squares sinusoidal fit as (T $\cdot \sqrt{sig}$)$^{-1}$ for a monoperiodic signal and (4T)$^{-1}$ if a second signal component is present within twice the classical frequency resolution defined by the time series length T. We are left with 59 frequencies ranging from 3.11$\mu$Hz to 93.7$\mu$Hz (see Tab.\,\ref{tab:obs}).

Figure\,\ref{FigFourier} illustrates our problem to safely determine intrinsic frequencies below approximately 20$\mu$Hz. The average amplitudes increase by a factor of about 3 and it is not clear -- nor can it be decided on the grounds of the present photometry -- which contribution comes from possibly yet undetected instrumental properties, granulation signal (1/f characteristics), pulsation, artefacts enforced by short mode lifetimes of genuine signal, or a combination of all this. Furthermore, {\it SigSpec} -- designed to handle white noise only -- cannot be applied anymore in this low frequency region. An unsatisfying work-around to take the increasing noise level towards the low frequencies into account could be to increase the acceptance level in the low frequency region from sig = 4 to about 36, as sig scales with power. We divide Tab.\,\ref{tab:obs} in a set of frequencies below 20$\mu$Hz, not used for modeling $\epsilon$ Oph, and above 20$\mu$Hz. 

We are then left with 39 frequencies (20 to 93.7$\mu$Hz). Because it is extraordinarily difficult to fit a model to a set of frequencies when the list of frequencies contains even a small number of spurious frequencies, i.e., of non stellar origin, we restrict our analysis to only those frequencies above 20$\mu$Hz with sig $\geq$ 10.0 (roughly corresponding to amplitudes $\geq$ 60 ppm). We are convinced that these large amplitude frequencies are truly intrinsic and not contaminated by instrumental or other effects (different from pulsation). There are 21 frequencies that satisfy these conditions, more than enough to constrain the models. The low frequency set was not used for modelling, but is listed for completeness. However, a surprinsingly large fraction of frequencies (8 of 19, indicated by an asterisk in Tab.\,\ref{tab:obs}) coincide with predictions of our best model (see Sec.\,4.1).

We searched the RV data for frequencies from 0 - 120$\mu$Hz ($\sim$10\d ) down to a sig limit of 5.2 (corresponding to S/N $\sim$ 3.9). We identify 25 frequencies. These are more ambiguous identifications because of the aliasing of the RV data; some of the frequencies may be $\pm$1 \d\ sidelobes of the real values. 11 of the 25 frequencies identified in the RV data are also present, within the frequency uncertainties, in the MOST photometric data. All significant RV frequencies are listed in Tab.\,\ref{tab:obs}.


   \begin{figure}[t]
   \centering
      \includegraphics[width=0.5\textwidth]{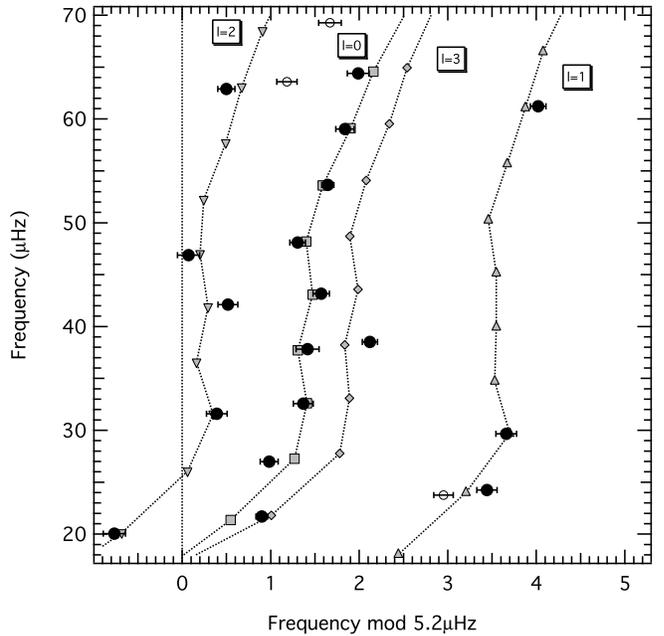}
      \caption{Echelle diagram of the 21 most significant observed oscillation frequencies (above 20$\mu$Hz) used to constrain the fit and the best fitting model modes. The observed photometric modes are shown as circles with error bars indicating observational uncertainties. Filled circles show to the 18 observed frequencies matched by model modes and open circles to the remaining 3 frequencies. The model modes are given by grey symbols connected by line segments. 
}
         \label{FigEchelle1}
   \end{figure}
   \begin{figure}[b]
   \centering
      \includegraphics[width=0.5\textwidth]{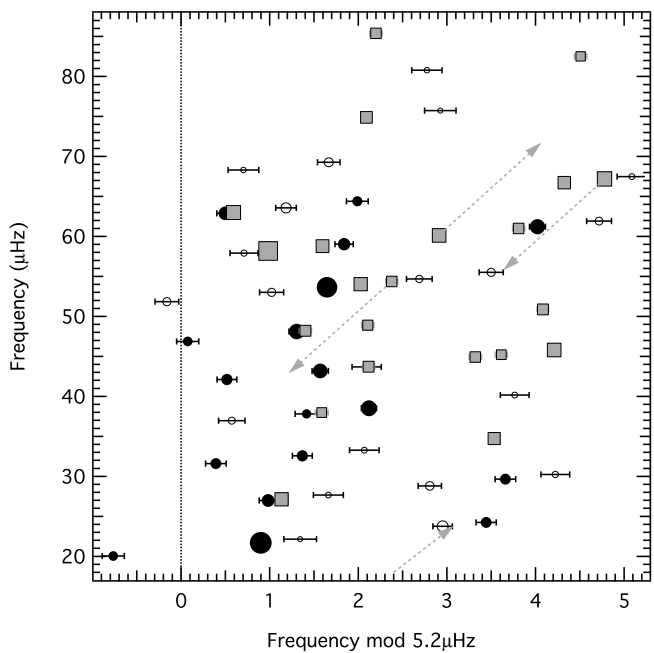}
      \caption{Same as Fig.\,\ref{FigEchelle1} but for all photometric (circles) and radial velocity (squares) frequencies. Symbol sizes are scaled with mode amplitudes: 40ppm to 125ppm and 0.9 to 3.4ms$^{-1}$ for the luminosity and radial velocity amplitudes, respectively. Arrows indicate RV modes which are presumably daily aliases.
}
         \label{FigEchelle2}
   \end{figure}

\section{Asteroseismic Analysis}
As a first step toward understanding the nature of the oscillations in $\epsilon$ Oph, we attempt to match the observed frequency spectrum with the eigenspectra determined from stellar models.  We calculated radial and nonradial pulsation frequencies of models distributed along evolutionary tracks on the giant branch of the HR Diagram. The models were constructed with the Yale Rotating Evolutionary Code (YREC, Guenther et al. 1992), adopting the OPAL equation of state (Iglesias \& Rogers 1996), OPAL opacities in the interior (Rogers \& Iglesias 1994) and the low-temperature opacities from Alexander \& Ferguson (1994) in the envelope. Diffusion of helium and heavy elements was not included. The model eigenspectra were generated by the nonradial, nonadiabatic stellar pulsation program developed by Guenther (1994). The program solves the linearised nonadiabatic oscillation equations using the Henyey relaxation method. 

The inert helium core in giant stars supports closely spaced $g$-modes at frequencies that overlap the frequency range of the $p$-modes. As a consequence there is significant mode mixing (also known as mode bumping) where the nonradial $p$-modes and $g$-modes interfere with each other (see Kallinger et al. 2007a and Guenther et al. 2000). Any nonradial $p$-modes seen on the surface of the star have $g$-mode character in the interior. For giants the radial order $n$ of the $g$-mode part of the mode can be very high ($\sim$20 to 100). These are the same modes identified as strongly trapped unstable (STU) modes by Dziembowski et al. (2001). The densely packed fluctuations in the radial displacement eigenfunctions near the stellar core heavily strain the numerical calculation and can lead to numerical instabilities that prevent convergence of a solution. To avoid this complication, we force the code to ignore the inner 10\% (in radius) of the stellar models. We have shown (Kallinger et al. 2007a) that the mode eigenfunctions in the outer 90\% of the model are practically the same if the inner part of the model is used or not. The disadvantage of this method is that one does not obtain the correct number of radial nodes for the $g$-mode part of the eigenfunction. For more details, see Kallinger et al. (2007a).

We computed radial and nonradial eigenfrequencies (with mode degree $l$ up to 3) for more than 37,000 models of intermediate-mass stars (from 1.5 to 3.5M$_{\sun}$ in steps of 0.02M$_{\sun}$) during their H-shell burning phase.  The model sequences start at temperatures around 5500K (near spectral type G1) and continue up to the initiation of He-core burning at the end of the red giant phase. We set the initial hydrogen and metal mass fractions to (X, Z) = (0.71, 0.01), to match the observationally determined metal abundance from Cayrel de Strobel et al. (2001), who found [Fe/H] = -0.25. The mixing length parameter, which adjusts the temperature gradient in convective regions, was set to $\alpha$ = 1.74 to match the findings of Yi et al. (2003). For the model physics we are using, a calibrated solar model would have $\alpha$ = 1.63. The value of $\alpha$ we adopt yields a model at the solar age and luminosity that is hotter than the Sun by $\sim$30K. 

The best fits between the observed frequency spectrum and our grid of model frequencies were found by searching for minima in $\chi^{2}$, defined as:
\vspace{0.3cm}
\begin{large}
\begin{center}
$\chi^{2} = \frac{1}{N} \sum _{i=1}^{N} \frac{(f_{obs,i} - f_{model,i})^{2}}{\sigma_{obs,i}^{2}}, $
\end{center}
\end{large}
\vspace{0.3cm}
where $f_{obs,i}$ and $f_{model,i}$ are the observed and corresponding model eigenfrequency of the ith mode, respectively, $\sigma_{obs,i}$ is the uncertainty of the observed frequency, and $N$ is the total number of modes used for the fit. This approach provides a reliable and unbiased estimate of how well a set of observed frequencies coincides with a model eigenspectrum. A value of $\chi^{2}$ $\leq$ 1.0 means that, on average, the model frequencies are within the uncertainty of the matched observed frequencies. For more details, see Guenther et al. (2005).  However, if there is more than one model with $\chi^{2}$ $\leq$ 1.0, the set of observed frequencies (and uncertainties) is not sufficient to identify a unique model. In this case, additional stellar parameters (such as independently determined temperature and/or luminosity) are necessary to narrow the parameter space of the model search.

\subsection{Mode Identifications}
Trying to match the observed oscillation spectrum to the models is made significantly more difficult if we are not sure that all the observed frequencies are $p$-modes intrinsic to the star. With 59 frequencies identified as possible modes it is computationally impractical to consider all possible combinations of $p$-modes versus spurious (i.e., not $p$-) modes. We therefor limited our model search to the 21 most significant frequencies (sig $>$ 10.0), as determined by {\it SigSpec}, in our list of 59 frequencies in the frequency range 20 to 95$\mu$Hz. 
The searching code will try to find the best fit to as many frequencies as possible and this may, as is the case here, require that some modes be excluded from the fit. The modes excluded varies from model to model in the search.  For the best fit model frequencies P23, P52, and P56 are unmatched. 
We were unable to find any models whose radial $p$-modes match more than 9 of the 21 observed frequencies. Only when we allowed the search to include nonradial modes up to and including $l$ = 3 were we able to fit 18 (bold in Tab.\,\ref{tab:obs}) of the 21 observed frequencies. The best model fit matches 9 of the frequencies to radial $p$-modes, 3 to $l$ = 1, 5 to $l$ = 2, and 1 to $l$ = 3 modes with a $\chi^{2}$ = 0.22. The best fitting model has a mass of 2.02 M$_{\sun}$, an effective temperature of 4892K and a luminosity of  60.1L$_{\sun}$, which places it within $\pm$1$\sigma$ of $\epsilon$ Oph's position in the HR-diagram. In Table 2 we list the adiabatic $p$-mode frequencies of the best fitting model up to the acoustic cut--off frequency of $\sim$92$\mu$Hz.

In Fig.\,\ref{FigEchelle1} we shown, in an echelle diagram with folding frequency 5.2$\mu$Hz, the 21 most significant photometric frequencies, used in the search for the best fitting model, along with the $p$-mode frequencies of the best model fit. The filled circles correspond to the 18 photometric frequencies that are fit by the model. The $p$-mode frequencies of the best fitting model are shown as small grey filled symbols connected by line segments.
In Fig.\,\ref{FigEchelle2} we show both the MOST (circles) and the RV (squares) frequencies to provide a direct comparison between the two data sets. Needless to say, there are some observed frequencies that do not appear to fall on the $l$-ridges of our best fit model. We argue below that some of these frequencies could still be intrinsic to the star.

\begin{table}[h]
\begin{center}
\caption{Oscillation frequencies (f), amplitudes (A) and {\it SigSpec} significances (sig) for $\epsilon$ Oph. Frequencies used to constrain the fit are marked in bold face. P23, P52 and P56 are also used in the model search procedure but do not match any mode of the best fitting model. ``+'' or ``-'' indicate RV frequencies which are presumably +1\,d$^{-1}$ or -1\,d$^{-1}$ aliases. Asterisk symbols indicate frequencies matched by modes of the best fitting model (see Tab.\,\ref{tab:model}). 
\label{tab:obs}}

\begin{tabular}{l | l l l |  l | l l l}
\hline
\hline
\noalign{\smallskip}
\multicolumn{8}{c}{Photometry}\\
\noalign{\smallskip}
\hline
\noalign{\smallskip}
P&f             &A     &sig &P&f             &A      &sig\\
  &{\tiny $\mu$Hz}&{\tiny ppm}&      &  &{\tiny $\mu$Hz}&{\tiny ppm}&      \\
\noalign{\smallskip}
\hline
\noalign{\smallskip}
{\tiny     01}		&{\tiny	    3.11$\pm$0.07} &{\tiny	 111 }&{\tiny	     30.3 }&	   {\tiny \bf 30$^{*}$}	&{\tiny  \bf 31.59$\pm$0.12} &{\tiny    \bf 69 }&{\tiny	 \bf 12.3 }\\
{\tiny     02}		&{\tiny	    3.50$\pm$0.15} &{\tiny	 54  }&{\tiny	      7.6 }&	   {\tiny \bf 31$^{*}$ }&{\tiny  \bf 32.57$\pm$0.11} &{\tiny   \bf 72  }&{\tiny    \bf 13.3 }\\
{\tiny     03$^{*}$}	&{\tiny	    3.92$\pm$0.06} &{\tiny	 146 }&{\tiny	     50.1 }&	   {\tiny     32$^{*}$ }&{\tiny      33.27$\pm$0.17} &{\tiny        48 }&{\tiny	  6.0 }\\
{\tiny     04}		&{\tiny	    5.98$\pm$0.09} &{\tiny	 96  }&{\tiny	     22.7 }&	   {\tiny     33$^{*}$ }&{\tiny      36.97$\pm$0.15} &{\tiny       53  }&{\tiny	  7.5 }\\
{\tiny     05$^{*}$}	&{\tiny	    6.44$\pm$0.04} &{\tiny	 190 }&{\tiny	     82.4 }&	   {\tiny \bf 34$^{*}$ }&{\tiny  \bf 37.82$\pm$0.13} &{\tiny    \bf 62 }&{\tiny    \bf 10.0 }\\
{\tiny     06$^{*}$}	&{\tiny	    7.54$\pm$0.03} &{\tiny	 254 }&{\tiny	    145.0 }&	   {\tiny \bf 35$^{*}$ }&{\tiny  \bf 38.52$\pm$0.09} &{\tiny   \bf 96  }&{\tiny    \bf 22.7 }\\
{\tiny     07$^{*}$}	&{\tiny	    8.49$\pm$0.06} &{\tiny	 152 }&{\tiny	     54.1 }&	   {\tiny     36$^{*}$ }&{\tiny      40.16$\pm$0.16} &{\tiny       48  }&{\tiny	  6.2 }\\
{\tiny     08}		&{\tiny	    9.07$\pm$0.07} &{\tiny	 121 }&{\tiny	     35.6 }&	   {\tiny \bf 37$^{*}$ }&{\tiny  \bf 42.12$\pm$0.11} &{\tiny   \bf 72  }&{\tiny    \bf 13.0 }\\
{\tiny     09}		&{\tiny	    9.84$\pm$0.05} &{\tiny	 176 }&{\tiny	     72.3 }&	   {\tiny \bf 38$^{*}$ }&{\tiny  \bf 43.17$\pm$0.09} &{\tiny   \bf 88  }&{\tiny    \bf 19.5 }\\
{\tiny     10$^{*}$}	&{\tiny	   11.83$\pm$0.07} &{\tiny	 126 }&{\tiny	     38.2 }&	   {\tiny     39$^{*}$ }&{\tiny      43.69$\pm$0.16} &{\tiny       48  }&{\tiny	  6.2 }\\
{\tiny     11}		&{\tiny	   12.83$\pm$0.06} &{\tiny	 145 }&{\tiny	     50.0 }&	   {\tiny \bf 40$^{*}$ }&{\tiny  \bf 46.87$\pm$0.13} &{\tiny    \bf 64 }&{\tiny    \bf 10.5 }\\
{\tiny     12$^{*}$}	&{\tiny	   13.99$\pm$0.09} &{\tiny	  88 }&{\tiny	     19.3 }&	   {\tiny \bf 41$^{*}$ }&{\tiny  \bf 48.11$\pm$0.09} &{\tiny    \bf 93 }&{\tiny    \bf 21.3 }\\
{\tiny     13$^{*}$}	&{\tiny	   15.20$\pm$0.16} &{\tiny	  49 }&{\tiny	      6.3 }&	   {\tiny     42$^{*}$ }&{\tiny      51.84$\pm$0.13} &{\tiny        59 }&{\tiny	  9.1 }\\
{\tiny     14$^{*}$}	&{\tiny	   15.99$\pm$0.06} &{\tiny	 142 }&{\tiny	     48.7 }&	   {\tiny     43 }&{\tiny      53.02$\pm$0.14} &{\tiny        59 }&{\tiny	  9.1 }\\
{\tiny     15}		&{\tiny	   16.29$\pm$0.13} &{\tiny	  64 }&{\tiny	     10.6 }&	   {\tiny \bf 44$^{*}$ }&{\tiny  \bf 53.65$\pm$0.07} &{\tiny   \bf 119 }&{\tiny    \bf 34.8 }\\
{\tiny     16}		&{\tiny	   16.65$\pm$0.06} &{\tiny	 131 }&{\tiny	     40.8 }&	   {\tiny     45 }&{\tiny      54.69$\pm$0.14} &{\tiny        55 }&{\tiny	  8.0 }\\
{\tiny     17}		&{\tiny	   17.50$\pm$0.08} &{\tiny	 109 }&{\tiny	     29.5 }&	   {\tiny     46$^{*}$ }&{\tiny      55.50$\pm$0.14} &{\tiny        59 }&{\tiny	  9.0 }\\
{\tiny     18}		&{\tiny	   19.04$\pm$0.12} &{\tiny	  69 }&{\tiny	     12.3 }&	   {\tiny     47$^{*}$ }&{\tiny      57.91$\pm$0.16} &{\tiny        50 }&{\tiny	  6.7 }\\
{\tiny     19}		&{\tiny	   19.48$\pm$0.06} &{\tiny	 146 }&{\tiny	     50.3 }&	   {\tiny \bf 48$^{*}$ }&{\tiny  \bf 59.04$\pm$0.10} &{\tiny    \bf 78 }&{\tiny    \bf 15.5 }\\
			&			   &		      &			   &	   {\tiny \bf 49$^{*}$ }&{\tiny  \bf 61.22$\pm$0.09} &{\tiny    \bf 92 }&{\tiny    \bf 20.9 }\\
{\tiny \bf 20$^{*}$}	&{\tiny  \bf 20.03$\pm$0.13} &{\tiny    \bf 64 }&{\tiny	 \bf 10.5 }&	   {\tiny     50 }&{\tiny      61.92$\pm$0.14} &{\tiny        57 }&{\tiny	  8.4 }\\
{\tiny \bf 21$^{*}$}	&{\tiny  \bf 21.70$\pm$0.07} &{\tiny   \bf 124 }&{\tiny	 \bf 37.3 }&	   {\tiny \bf 51$^{*}$ }&{\tiny  \bf 62.90$\pm$0.10} &{\tiny    \bf 83 }&{\tiny    \bf 17.5 }\\
{\tiny     22$^{*}$}	&{\tiny	   22.14$\pm$0.18} &{\tiny	  43 }&{\tiny	      4.9 }&	   {\tiny     52 }&{\tiny      63.58$\pm$0.12} &{\tiny        70 }&{\tiny	 12.5 }\\
{\tiny     23}		&{\tiny	   23.75$\pm$0.11} &{\tiny	  73 }&{\tiny	     13.5 }&	   {\tiny \bf 53$^{*}$ }&{\tiny  \bf 64.39$\pm$0.12} &{\tiny    \bf 65 }&{\tiny    \bf 10.9 }\\
{\tiny \bf 24$^{*}$}	&{\tiny  \bf 24.24$\pm$0.12} &{\tiny    \bf 70 }&{\tiny	 \bf 12.5 }&	   {\tiny     54 }&{\tiny      67.49$\pm$0.17} &{\tiny        48 }&{\tiny	  6.0 }\\
{\tiny \bf 25$^{*}$}	&{\tiny  \bf 26.98$\pm$0.10} &{\tiny    \bf 80 }&{\tiny	 \bf 16.3 }&	   {\tiny     55$^{*}$ }&{\tiny      68.30$\pm$0.17} &{\tiny        46 }&{\tiny	  5.6 }\\
{\tiny     26$^{*}$}	&{\tiny	   27.66$\pm$0.17} &{\tiny	  47 }&{\tiny	      5.9 }&	   {\tiny     56 }&{\tiny      69.27$\pm$0.13} &{\tiny        63 }&{\tiny	 10.2 }\\
{\tiny     27}		&{\tiny	   28.81$\pm$0.13} &{\tiny	  61 }&{\tiny	      9.6 }&	   {\tiny     57$^{*}$ }&{\tiny      75.72$\pm$0.18} &{\tiny        44 }&{\tiny	  5.2 }\\
{\tiny \bf 28$^{*}$}	&{\tiny  \bf 29.66$\pm$0.12} &{\tiny    \bf 69 }&{\tiny	 \bf 12.3 }&	   {\tiny     58 }&{\tiny      80.77$\pm$0.17} &{\tiny        46 }&{\tiny	  5.7 }\\
{\tiny     29}		&{\tiny	   30.22$\pm$0.16} &{\tiny	  49 }&{\tiny	      6.3 }&	   {\tiny     59 }&{\tiny      93.74$\pm$0.18} &{\tiny        44 }&{\tiny	  5.1 }\\

\noalign{\smallskip}
\hline
\noalign{\smallskip}
\multicolumn{8}{c}{Radial velocity}\\
\noalign{\smallskip}
\hline
\noalign{\smallskip}
R&f             &A              &sig &R&f             &A              &sig\\
  &{\tiny $\mu$Hz}&{\tiny m/s}&      &  &{\tiny $\mu$Hz}&{\tiny m/s}&      \\
\noalign{\smallskip}
\hline
\noalign{\smallskip}
{\tiny 01$^{*}$}&{\tiny  11.55$\pm$0.06} &{\tiny 2.1 }&{\tiny   5.8 } &   {\tiny 14$^{*}$}&{\tiny  54.03$\pm$0.05} &{\tiny 2.0 }&{\tiny  10.2 } \\
{\tiny 02$_{-}^{*}$}&{\tiny  12.31$\pm$0.07} &{\tiny 0.9 }&{\tiny   5.4 } &   {\tiny 15$_{+}^{*}$}&{\tiny  54.38$\pm$0.06} &{\tiny 1.2 }&{\tiny   5.8 } \\
{\tiny 03 }&{\tiny  16.86$\pm$0.07} &{\tiny 1.1 }&{\tiny   5.2 } &   {\tiny 16$^{*}$}&{\tiny  58.18$\pm$0.03} &{\tiny 3.4 }&{\tiny  21.2 } \\
{\tiny 04$^{*}$}&{\tiny  27.13$\pm$0.05} &{\tiny 1.9 }&{\tiny   9.0 } &   {\tiny 17$^{*}$}&{\tiny  58.80$\pm$0.05} &{\tiny 1.8 }&{\tiny   9.3 } \\
{\tiny 05$^{*}$}&{\tiny  34.73$\pm$0.06} &{\tiny 1.6 }&{\tiny   7.8 } &   {\tiny 18$_{-}^{*}$}&{\tiny  60.11$\pm$0.05} &{\tiny 2.1 }&{\tiny   9.8 } \\
{\tiny 06$^{*}$}&{\tiny  37.99$\pm$0.07} &{\tiny 1.0 }&{\tiny   5.2 } &   {\tiny 19$^{*}$}&{\tiny  61.01$\pm$0.06} &{\tiny 1.2 }&{\tiny   5.8 } \\
{\tiny 07$^{*}$}&{\tiny  43.72$\pm$0.06} &{\tiny 1.3 }&{\tiny   5.8 } &   {\tiny 20$^{*}$}&{\tiny  62.99$\pm$0.05} &{\tiny 2.1 }&{\tiny   9.7 } \\
{\tiny 08 }&{\tiny  44.92$\pm$0.07} &{\tiny 1.2 }&{\tiny   5.5 } &   {\tiny 21$^{*}$}&{\tiny  66.72$\pm$0.05} &{\tiny 1.7 }&{\tiny   9.0 } \\
{\tiny 09$^{*}$}&{\tiny  45.21$\pm$0.06} &{\tiny 1.0 }&{\tiny   5.8 } &   {\tiny 22$_{+}^{*}$}&{\tiny  67.18$\pm$0.05} &{\tiny 2.2 }&{\tiny   9.9 } \\
{\tiny 10 }&{\tiny  45.81$\pm$0.05} &{\tiny 2.1 }&{\tiny   9.8 } &   {\tiny 23 }&{\tiny  74.89$\pm$0.06} &{\tiny 1.5 }&{\tiny   7.0 } \\
{\tiny 11$^{*}$}&{\tiny  48.20$\pm$0.06} &{\tiny 1.4 }&{\tiny   6.1 } &   {\tiny 24 }&{\tiny  82.51$\pm$0.07} &{\tiny 0.9 }&{\tiny   5.5 } \\
{\tiny 12$^{*}$}&{\tiny  48.91$\pm$0.07} &{\tiny 1.1 }&{\tiny   5.4 } &   {\tiny 25$^{*}$}&{\tiny  85.40$\pm$0.07} &{\tiny 1.2 }&{\tiny   5.5 } \\
{\tiny 13 }&{\tiny  50.88$\pm$0.06} &{\tiny 1.3 }&{\tiny   6.4 } &&&&\\

\noalign{\smallskip}
\hline

\end{tabular}
\end{center}
\end{table}

\begin{table}[h]
\begin{center}
\caption{Adiabatic $p$-mode frequencies of the best fitting model up to the acoustic cut--off frequency in units of $\mu$Hz. For nonradial modes $n_{p}$ is the number of radial nodes in the $p$--mode part of the modes. Identifier of observed modes matching a model frequency are listed in brackets. Bold faced modes indicate the modes matched by observed frequencies used for model fitting.
\label{tab:model}}

\begin{tabular}{lllll}
\hline
\hline
\noalign{\smallskip}
  $n_{p}$	&	 \multicolumn{1}{c}{$l$=0}	&	\multicolumn{1}{c}{$l$=1}	&	  \multicolumn{1}{c}{$l$=2}	    &	     \multicolumn{1}{c}{$l$=3}	\\
\noalign{\smallskip}
\hline
\noalign{\smallskip}
  0	&    \multicolumn{1}{r}{7.974  $P06 \choose  -$}          &    \multicolumn{1}{r}{3.668  $P03 \choose  -$}           &     \multicolumn{1}{r}{6.797  $P05 \choose  -$}            &    \multicolumn{1}{r}{8.416  $P07 \choose  -$}        \\
\noalign{\smallskip}
  1	&   15.170 $ P13 \choose  -$           &   11.540 $P10 \choose  R01$     &    13.856 $P12 \choose  -$            &   15.561 $P14 \choose  -$        \\
\noalign{\smallskip}
  2	&   \bf 21.348 $ P21 \choose  -$       &   18.042                 &    \bf 20.119 $P20 \choose  -$        &   21.807 $P22 \choose  -$        \\
\noalign{\smallskip}
  3	&   \bf 27.268 $P25 \choose  R04$ &   \bf 24.005 $P24 \choose  R02$       &    26.062 	            &	27.779 $P26 \choose  -$	       \\
\noalign{\smallskip}
  4	&   \bf 32.609 $P31 \choose  -$       &   \bf 29.700 $P28 \choose  -$       &    \bf 31.549 $P30 \choose  -$        &	33.089 $P32 \choose  -$	       \\
\noalign{\smallskip}
  5	&   \bf 37.706 $P34 \choose  R06$ &   34.731 $-  \choose R05$	    &    36.562 $P33 \choose  -$ 	    &   \bf 38.235 $P35 \choose  -$    \\
\noalign{\smallskip}
  6	&   \bf 43.073 $P38 \choose  R15$ &   39.948 $P36 \choose  -$	    &    \bf 41.892 $P37 \choose  -$        &   43.585 $P39 \choose  R07$  \\
\noalign{\smallskip}
  7	&   \bf 48.200 $ P41 \choose  R11$ &   45.149 $-  \choose  R09$	    &    \bf 47.002 $P40 \choose  -$        &   48.694 $-  \choose  R12$        \\
\noalign{\smallskip}
  8	&   \bf 53.588 $ P44 \choose  -$       &   50.257 	            &    52.243 $P42 \choose  -$ 	    &   54.078 $-  \choose  R14$        \\
\noalign{\smallskip}
  9	&   \bf 59.104 $P48 \choose  R17$ &   55.671 $P46 \choose  R22$	    &    57.693 $P47 \choose  R16$ 	    &   59.539           \\
\noalign{\smallskip}
 10	&   \bf 64.560 $P53 \choose  -$       &   \bf 61.077 $P49 \choose  R19$ &    \bf 63.074 $P51 \choose  R20$  &	64.940 	       \\
\noalign{\smallskip}
 11	&   70.117                    &   66.475 $-  \choose  R21$           &    68.508 $P55 \choose  -$            &   70.434           \\
\noalign{\smallskip}
 12	&   75.747 $P57 \choose  -$           &   71.991 $-  \choose  R18$           &    74.035                     &   75.985           \\
\noalign{\smallskip}
 13	&   81.366                    &   77.539                     &    79.559                     &   81.516          \\
\noalign{\smallskip}
 14	&   87.004                    &   83.083                     &    85.086 $-  \choose  R25$            &   87.061          \\
\noalign{\smallskip}
 15	&   92.639                    &   88.652                     &    90.625                      &   92.597           \\

\noalign{\smallskip}
\hline
\end{tabular}
\end{center}
\end{table}
The two frequencies P05 and P06 (6.44 and 7.54$\mu$Hz) that have the largest photometric amplitudes, but lie outside the 20 to 95 $\mu$Hz range used to constrain the best fit model, do not match (within the uncertainties) any of the low $n$-valued $p$-modes of the best fit model but are still close to the fundamental mode of the radial and $l$=2 sequence. If these two frequencies are indeed low order $p$-modes than they imply that the deep interior structure of our models is slightly off. This is possible since the higher order modes mainly constrain the outer layers of the model. It is possible that a different assumed initial helium abundance for the model, which would affect the frequencies of very lowest order modes, could yield a model that fits the lowest two frequencies along with the rest of the frequencies. We will address this and other modeling interpretations of the original 59 MOST frequencies in a separate paper.

Nine of the frequencies detected in the RV data match the $p$-mode frequencies of the best fitting model and are found in the list of photometrically observed frequencies, and 6 of the RV frequencies match the best fitting model and are not found in the list of  photometrically observed frequencies. We note that 4 RV frequencies would lie close to model frequencies if they are 1 cycle/day aliases of the real frequencies. The presumed real frequencies are at the tips of the arrows in Fig.\,\ref{FigEchelle2}. 

Although we are able to fit 18 of the 21 photometric frequencies with the largest {\it SigSpec} significances, only 63\% (``*'' in Tab.\,\ref{tab:obs}) of the photometric frequencies (70\% in the frequency region above 20$\mu$Hz) and 76\% of the RV frequencies can be attributed to pulsation modes in our best-fitting model. A few of the lowest frequencies may be associated with turbulence or rotational modulation in the envelope of the red giant or, as noted above, may not fit the model because the model does not have the correct helium abundance. We speculate, though, that frequencies that are not matched by the model could be the result of the damping and stochastical re-excitation of the modes. These modulations introduce spurious peaks around the true frequencies of modes in Fourier space. We discuss this hypothesis in Section 4.4 below.  

We note that Hekker et al. (2006) attempted mode identifications for the dominant frequency in the $\epsilon$ Oph spectroscopic data (De Ridder et al. 2006). They identified the frequency R16 at 58.2$\mu$Hz (having the largest RV amplitude) to be a mode of ($l, m$) = ($2, 2$). Our best fitting model matches this frequency with a mode of degree $l$ = 2 as well.


   \begin{figure}[h]
   \centering
      \includegraphics[width=0.5\textwidth]{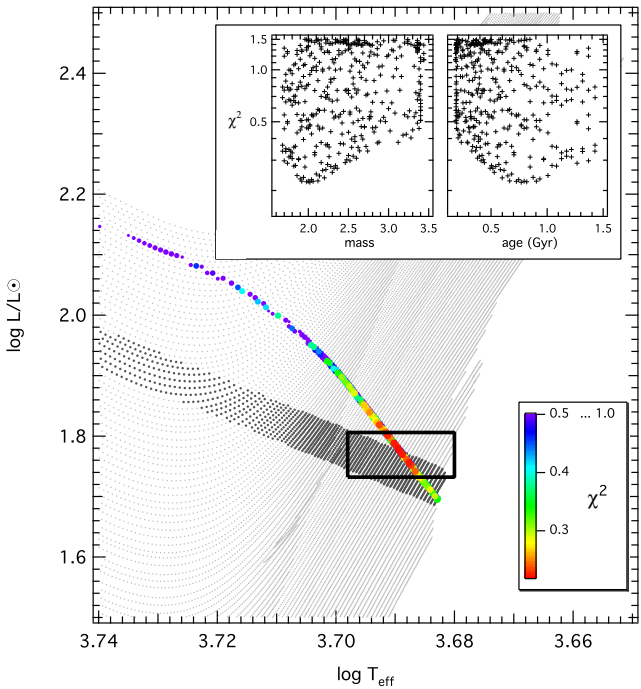}
      \caption{Theoretical HR--Diagram showing the uncertainty box location of $\epsilon$ Oph and a subset of the stellar model grid used for the pulsation analysis (light grey dots). The color scale gives the $\chi^{2}$ values from the differences between the observed and model frequencies, where the scale is limited to values less than 1.0 (for better visibility the color dark blue includes all models with $\chi^{2}$ from 0.5 to 1.0). $\chi^{2} <$ 1 indicated that the model frequencies are within 1$\sigma$ of the observed frequency uncertainty. The dark grey dots show models with a stellar radius of 10.4$\pm$0.45R$_{\sun}$ . The insert illustrates $\chi^{2}$ as a function of model mass and age, respectively.}
         \label{FigHRD}
   \end{figure}

\subsection{Position of $\epsilon$ Oph in the HR--Diagram}
A summary of fundamental parameters of the best fitting model and values found in the literature is given in Table 2.
De Ridder et al. (2006) conducted an extensive literature search to locate $\epsilon$ Oph reliably in the HR--Diagram. They adopted an effective temperature of 4887$\pm$100K and a luminosity of 59$\pm$5L$_{\sun}$ . These values compare well with our best fit model which has a surface temperature of 4892K and a luminosity of 60.1L$_{\sun}$. 

In Fig.\,\ref{FigHRD}, we present a theoretical HR--Diagram showing a subset of the stellar model grid used for the pulsation analysis (small light grey points), the best model fits to the observed frequencies with $\chi^{2}$ less than 1.0 (color), and the error box for the luminosity and effective temperature of $\epsilon$ Oph. The models with the smallest deviation from the oscillation observations (0.28$< \chi^{2} < $0.3) are shown in red, and lie close to the centre of the error box. We remind the reader that $\chi^{2}$ is determined from the oscillation data only, that is, other observables are not used in any way to help determine the best model fit. The insets in Fig.\,\ref{FigHRD} show $\chi^{2}$ as a function of model mass and age. They illustrate the smooth distribution of $\chi^{2}$ along the track of best fitting models (colored region) with constant large frequency separation.

Richichi et al. (2005) list an angular diameter of 2.94$\pm$0.08mas for $\epsilon$ Oph in their catalogue of stellar radii obtained with the ESO VLTI. This angular diameter and the Hipparcos parallax of 30.34$\pm$0.79mas give a stellar radius of 10.4$\pm$0.45R$_{\sun}$ . In Fig.\,\ref{FigHRD}, all models with a stellar radius within that range are marked by dark grey dots. The radius of the best-fitting model is about 10.8R$_{\sun}$ , within the uncertainty of the interferometrically determined radius.

\subsection{Mode Amplitudes}
Using the effective temperature, mass and radius of our best-fitting model of $\epsilon$ Oph (see Section 4.2), we obtain from the Kjeldsen \& Bedding (1995) scaling relation a radial order of $n$ = 9 at a frequency of $\sim$59$\mu$Hz for the mode of maximum power. The frequency with the largest RV amplitude is R14 at 58.2$\mu$Hz, corresponding to a mode with radial order $n$ = 9 in our best fitting model and consistent with Kjeldsen and Bedding's scaling relation. 

Kjeldsen \& Bedding (1995) provide a heuristically determined relation between luminosity and RV amplitudes for solar-like oscillations scaled by the amplitude ratio of solar $p$-modes in light and velocity: 20.1 ppm per ms$^{-1}$ at a wavelength of 550 nm. 

The average amplitude of the frequencies in the photometry of $\epsilon$ Oph matched to $p$-modes in the model (plus the two largest-amplitude peaks) is 85ppm.  The corresponding value for the RV data is 1.64ms$^{-1}$.  This corresponds to an amplitude ratio of 51.8ppm per ms$^{-1}$.  If we perform a linear regression of the photometric and RV amplitudes for the subset of frequencies identified in both data sets, we obtain a slope of 41$\pm$7ppm per ms$^{-1}$ (we must neglect the frequency $\sim$58$\mu$Hz with the largest RV amplitude to obtain this fit). This slope results in the following relation calibrated by the luminosity and temperature of $\epsilon$ Oph and the central wavelength of the MOST passband:
\vspace{0.3cm}
\begin{large}
\begin{center}
$(\delta L/L)_{\lambda} = \frac{v_{osc}/ms^{-1}}{(\lambda/550nm)(T_{eff}/5777K)^2}26.4\pm4.6ppm, $
\end{center}
\end{large}
\vspace{0.3cm}
The coefficient (26.4$\pm$4.6 ppm per ms$^{-1}$) is close to that of the Kjeldsen \& Bedding (1995) amplitude ratio for solar $p$-modes (20.1 ppm per ms$^{-1}$).  However, we caution that the reduction of the $\epsilon$ Oph spectroscopic time series was performed with respect to a nightly reference spectrum, which does suppress lower-frequency power and likely means a decrease of the measured RV amplitudes with increasing frequency relative to the intrinsic values.

\subsection{Mode Lifetimes}
Unlike $p$-modes in stars in the classical instability strip, which can coherently oscillate for millions of years, solar-type oscillations driven stochastically by turbulence in the convective envelope have much shorter lifetimes. Pulsations of this nature can be characterized by a superposition of randomly excited, intrinsically damped, harmonic oscillations. The Fourier transform of a set of such incoherent signals produces peaks randomly distributed under a Lorentzian profile with a width given by the damping rate (inversely proportional to the mode lifetime). 

In this work, we treat the individual frequencies in the data as independent radial and nonradial modes whose lifetimes are comparable to the duration of the observations.  For solar-type oscillations, the scatter of observed frequencies about the true mode frequencies is a function of the mode lifetime (and the signal-to-noise ratio of the data).  Kjeldsen et al. (2005) performed extensive simulations in their analysis of the RV oscillations of $\alpha$ Cen B to quantify the dependence of the frequency scatter on the lifetime and the S/N ratio. 

\begin{figure}[t]
   \centering
      \includegraphics[width=0.5\textwidth]{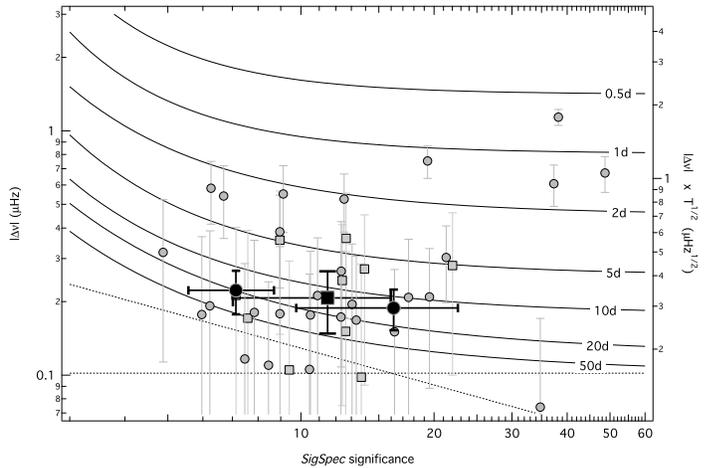}
      \caption{Calibration of mode lifetimes for the observed oscillation modes. As a result of numerical simulations of solar-type oscillations, solid lines indicate how much a detected frequency deviates on average from the input frequency as a function of the {\it SigSpec} significance for various mode lifetimes. The left axis gives the absolute deviation $| \Delta \nu |$ for the $\epsilon$ Oph observing window and to be independent from the data set length T, the right axis gives the absolute deviation normalized to T$^{1/2}$. The dotted lines correspond to the upper limit of the observational frequency uncertainty of a stable sinusoidal multi--periodic (horizontal line) and a mono--periodic signal (tilted line), respectively. Both are a function of T and are only valid for the left axis. Squares indicate the absolute deviations between frequencies determined from MOST and RV data for the same modes with an average value given by the large square symbol. Dots label the absolute deviation of all photometric mode frequencies from that expected from the neighboring modes with equal degree $l$, using linear interpolation. Average values are shown as large dots for two ranges of {\it SigSpec} significance (below and above 10).}
         \label{FigLifetime}
   \end{figure}

We apply a similar approach, assessing roughly 100,000 simulated data sets using the time sampling of the MOST $\epsilon$ Oph photometry.  Each data set contains a damped and randomly re-excited oscillation at a given input frequency, generated by the method described by Chaplin et al. (1997), plus white noise of random amplitude. The generated mode lifetimes range from 0.5 to 50 days. The average absolute deviation of the highest amplitude peak (determined by {\it SigSpec}) from the input frequency, $\Delta \nu$, as a function of the peak's {\it SigSpec} significance is presented in Fig.\,\ref{FigLifetime} for different mode lifetimes (solid lines). The left axis scale in this figure is specific to the MOST data run. From additional simulations with data sets of different length T we found that the frequency deviation due to finite mode lifetime scales with the square root of T. The right--hand axis in Fig.\,\ref{FigLifetime} is calibrated to be independent of T.  We also present upper limits for the observational uncertainties of a stable mono--periodic signal (horizontal dashed line scaled to the left axis) and a multi--periodic signal (diagonal dashed line, also, scaled to the left axis). (For more background details, see Kallinger et al. 2007b). 

As an example, consider some modes in our MOST observations of $\epsilon$ Oph with {\it SigSpec} significance of 10 or more. If the mode lifetimes are 1 day on average, then according to our simulation we can expect the observed frequencies of the modes to be on average within 1$\mu$Hz of the intrinsic frequencies of the modes. If the modes have a lifetime of 20 days, then our simulation indicates that the observed frequency of the modes will be on average within 0.2$\mu$Hz of the intrinsic frequencies of the modes.

\begin{table}[t]
\begin{center}
\caption{Fundamental stellar parameters for $\epsilon$ Oph as found in the literature and for the best fitting model. 
\label{tab2}}

\begin{tabular}{lcc}
\hline
\hline
\noalign{\smallskip}
 & Literature & best fitting \\
 &&model\\
\noalign{\smallskip}
\hline
\noalign{\smallskip}
Effective temperature  ...  [K]		&4877$\pm$100$^1$ &4892\\
Luminosity  ...  [L$_{\sun}$]	&59$\pm$5$^1$	&60.13\\
Radius  ...  [R$_{\sun}$]	&10.4$\pm$0.45$^2$&10.82\\
Mass  ...  [M$_{\sun}$]	&				&2.02\\
log g  ...  [g cm s$^{-2}$]	&2.48$\pm$0.36$^3$&2.674\\
Age  ...  [Gyr]		&				&0.770\\
Mixing length parameter  ...  [Hp]		&				&1.74\\
Metallicity  ...  [Z]		&0.01$^4$		&0.01\\
\noalign{\smallskip}
\hline
\noalign{\smallskip}
\multicolumn{3}{l}{$^1$De Ridder et al. (2006)}\\
\multicolumn{3}{l}{$^2$Richichi et al. (2005) and Hipparcos}\\
\multicolumn{3}{l}{$^3$Allende \& Lambert (1999)}\\
\multicolumn{3}{l}{$^4$[Fe/H] = -0.25 from Cayrel de Strobel et al. (2001)}\\
\end{tabular}
\end{center}
\end{table}

We can invert this, by using the apparent scatter of observed frequencies to estimate the mode lifetimes. To determine the scatter in the observed frequencies we can either compare the frequencies of the modes observed in RV with the frequencies of the same modes observed in MOST photometry or we can compare the observed MOST frequencies with the nearest corresponding frequency in the best fit model. Unfortunately, the latter method depends on the model fit itself. To avoid this we adopt the approach of Kjeldsen et al. (2005). We assume the modes to lie along ridges of common $l$--value and take the amount the modes deviate from the ridge as the scatter frequency (i.e., frequency difference). The frequency difference is taken to be the amount the mode's frequency deviates from a line drawn between the mode below and the mode above the mode of common $l$--value. Clearly, both of these methods depend on the correct identification of the modes. We present the results of both methods. 

In Fig.\,\ref{FigLifetime} we plot, using light grey squares, the frequency differences (i.e., scatter) between the 12 modes common in the RV and MOST determined frequencies as a function of their {\it SigSpec} significance. The data points all (except one) lie below the 5d lifetime line of our simulation. Indeed, the average value of the frequency differences, shown as a black square in Fig.\,\ref{FigLifetime}, corresponds to an average mode lifetime of $\sim$12 days. 

Also in Fig.\,\ref{FigLifetime} we plot, using light grey circles, the frequency differences of the MOST photometric data assuming the intrinsic mode frequencies do lie along smooth curves of common l-value. Although the scatter is greater than using the previous method, the averaged absolute deviations in the {\it SigSpec} significance range below and above 10 correspond to $\sim$20 day and $\sim$13 day mode lifetimes, respectively (black circles in Fig.\,\ref{FigLifetime}). The light grey circles in the upper right corner of Fig.\,\ref{FigLifetime}, correspond to low frequency modes, i.e., modes that naturally deviate the most from the regular spacing of the asymptotic formula. Both methods give similar average frequency deviations allocated to mode lifetimes of approximately 10 to 20 days.

We also performed simulations to estimate the likelihood of a second (``spurious'') peak appearing in Fourier space, given a mode with a lifetime 10--20 days and {\it SigSpec} significances comparable to those observed in $\epsilon$ Oph. The simulation assumed the time sample cadence of the MOST photometry and, separately, the time sample cadence of the RV data. We find that almost half of the frequencies will appear as ``spurious'' and apparently do not match the model, as we have found in Section 4.1 when considering all 59 photometric frequencies.


\section{Discussion \& Conclusions}
By demonstrating the mutual consistency between our best fit models and the observations we conclude that observable nonradial modes do exist in giants. 

We have shown that the significant intrinsic frequencies present in the MOST photometry (Barban et al. 2007) and CORALIE + ELODIE RV measurements (De Ridder et al. 2006) of $\epsilon$ Oph can be best matched to a model of radial and nonradial $p$-modes.  The number of matched $p$-modes compared to unmatched modes as well as the scatter of the observed frequencies around the mode frequencies is consistent with mode lifetimes of 10--20 days.

Like Barban et al. (2007), we find in the MOST data a clear well-populated sequence of radial modes with a fundamental spacing of $\sim$ 5.8$\mu$Hz in the asymptotic regime and $\sim$ 5.2$\mu$Hz in the range where most of the frequencies were observed. But for the 21 most significant frequencies in the photometric data above 20$\mu$Hz, we are only able to fit 9 of the frequencies to radial $p$-modes. Furthermore, we cannot find any model in our grid of more than 30,000 models that matches all 7 radial mode frequencies found by Barban et al. (2007) via Lorentzian profile fitting. We can, though, match 18 of the 21 most significant frequencies in the photometric data if we consider nonradial modes. 

The $l$ = 0 sequence is nearly fully populated and the $l$ = 2 sequence is also well populated, while the $l$ = 1 and 3 sequences are less densely populated. This is not so surprising for $l$ = 3 modes because their photometric and RV amplitudes in light integrated over the stellar disk are expected to be very small, even given the high precision of the data. But from the geometrical aspects of the spherical harmonics, $l$ = 1 modes should have larger amplitudes than $l$ = 2 modes for almost every orientation except nearly pulsational--equator--on. A possible explanation is suggested by Dziembowski et al. (2001). Their theoretical models show that $l$ = 2 nonradial modes are more frequently unstable than others, hence, $l$ = 1 modes are less likely to be observable. In general, the distribution of modes according to $l$--value for $\epsilon$ Oph resembles what Kallinger et al. (2007a) found for the red giant HD 20884.

The intrinsic nature of the most significant observed frequencies is also supported by the good agreement between the oscillation-only constrained best-fit model and the independently derived effective temperature, luminosity, and radius. Our best fit model has an effective temperature of 4892K and luminosity of 60.1L$_{\sun}$ compared to, e.g., De Ridder et al. (2006) non-asteroseismically derived temperature 4887$\pm$100K and luminosity 59$\pm$5L$_{\sun}$. Furthermore our best model fit has a radius of 10.82R$_{\sun}$ that agrees with the radius 10.4$\pm$0.45R$_{\sun}$ based on ESO VLTI angular diameter determination of Richichi et al. (2005) and the Hipparcos parallax. The mass and age of $\epsilon$ Oph as constrained by the oscillations is 2.02M$_{\sun}$ and 0.77Gyr.

\begin{acknowledgements}
The Natural Sciences and Engineering Research Council of Canada supports the research of D.B.G., J.M.M., A.F.J.M., S.M.R., and A.F.J.M. is also supported by FQRNT (Qu\'ebec), and R.K. is supported by the Canadian Space Agency. T.K, D.H. and W.W.W. are supported by the Austrian Research Promotion Agency (FFG), and the Austrian Science Fund (FWF P17580). Furthermore, it is a pleasure to thank C. Aerts and J. De Ridder (Universiteit Leuven) for providing us with the radial velocity data.
\end{acknowledgements}

\end{document}